\documentclass[useAMS,usenatbib]{mn2e}

\newcommand{\dg}{$^\circ$ }
\newcommand{\kms}{km~s$^{-1}$ }
\usepackage{graphicx}

\title[Satellite Galaxy Orbits]{The Orbital Distribution of Satellite Galaxies}
\author[S.\ Herbert-Fort et al.]{St\'ephane Herbert-Fort$^{1}$,
Dennis Zaritsky$^{1}$,
Yeun Jin Kim$^{2,1}$,
Jeremy Bailin$^{3,4}$, 
\newauthor
and
James E.\ Taylor$^{5}$
\vspace{0.3cm}
\\
$^1$University of Arizona/Steward Observatory, 933 N Cherry Avenue, 
Tucson, AZ 85721 (email: shf@as.arizona.edu)\\
$^2$Department of Astronomy and Astrophysics, University of Chicago,
5640 S. Ellis Avenue, Chicago, IL  60637\\
$^3$Department of Physics \& Astronomy, ABB-241,
McMaster University, Hamilton, ON, L8S4M1, Canada\\
$^4$Centre for Astrophysics \& Supercomputing, Swinburne University, 
Hawthorn, VIC 3122, Australia\\
$^5$Department of Physics \& Astronomy, University of Waterloo, 
Waterloo, ON, N2L3G1, Canada}
\begin{document}

\date{Accepted 2007 November 21}

\pagerange{\pageref{firstpage}--\pageref{lastpage}} \pubyear{2007}

\maketitle

\label{firstpage}

\begin{abstract}
We measure the distribution of velocities for prograde and retrograde 
satellite galaxies using a combination of published data and new observations 
for 78 satellites of 63 extremely isolated disc galaxies (169 satellites total). 
We find that the velocity distribution 
is non-Gaussian ($>$ 99.9\% confidence), but that it can be described as the sum of
two Gaussians, one of which is
broad ($\sigma = 176 \pm 15$ km s$^{-1}$), has a mean prograde 
velocity of  $86\pm30$ km s$^{-1}$, and 
contains $\sim$ 55\% of the satellites, while the other
is slightly retrograde with a mean velocity of $-21 \pm 22$ km s$^{-1}$ 
and $\sigma = 74 \pm 18$ km s$^{-1}$ 
and contains $\sim$ 45\% of the satellites. Both of these components are present 
over all projected radii and found in the sample regardless of cuts 
on primary inclination or satellite disc angle.  The double-Gaussian 
shape, however, becomes more pronounced among satellites of more 
luminous primaries.  We remove the potential dependence of satellite velocity on
primary luminosity using the  
Tully-Fisher relation and still find the velocity distribution to be asymmetric and
even more significantly non-Gaussian.  
The asymmetric velocity distribution demonstrates
a connection between the inner, visible disc galaxy and the kinematics 
of the outer, dark halo. The reach of this connection, extending even beyond the 
virial radii, suggests that it is imprinted by the satellite infall pattern 
and large-scale effects, rather than by higher-level dynamical 
processes in the formation of the central galaxy or late-term evolution of
the satellites.
\end{abstract}


\section{Introduction}
Although it is now generally accepted that galaxies are embedded in large ($>$100
kpc), massive ($>10^{12} M_\odot$) dark matter haloes, 
the detailed spatial and kinematic structure
of those haloes, and the relationships between that structure and the resulting optical 
galaxy, remain largely unknown.  The particulars of the process of accretion and 
assembly are usually deduced from simulations because the inner galaxy, which is
what we can easily observe, has been scrambled by its complex physical evolution. 
We seek empirical evidence, in the outer dark halo, of how disc galaxies grow 
and signatures of a connection between the outer halo and the inner galaxy.
The only tracers that are sufficiently 
luminous and distant from the central galaxy are satellite galaxies (Zaritsky 
et al. 1993; Zaritsky et al. 1997, hereafter ZSFW).  

Two relatively recent advances justify a re-examination of this topic.
First, the sample of spectroscopically confirmed satellites has 
grown by more than an order of magnitude recently due to the Two-Degree Field 
Galaxy Redshift Survey (2dFGRS; Sales \& Lambas 2004, 2005) 
and the Sloan Digital Sky Survey \citep[SDSS;][]{Prada03}.  Second, a slew of new 
simulations is aimed at explaining/predicting the properties of satellite galaxies 
\citep[e.g.][]{Knebe04, vdB05, Zent05, Azzaro06, Sharma05, L07, Sales07}.  
Overall, simulations agree that infalling material is critical to the 
formation of galaxies and that satellites should bear some
imprint of this infall. Some naive scenarios suggest potential
signatures.  First, for a homogeneous collapse model of 
galaxy formation (in which the satellites are unbiased fragmented 
debris of the process; e.g.\ Maller \& Bullock 2004), one 
would predict a high degree of alignment between 
the disc and satellite angular momenta because
both arise from the same parent population.
In contrast, in a hierarchical model,
one might expect late-coming satellites, those preferentially found 
at large radii, to have little connection to the central disc. To complicate matters further, 
in a model with strong dynamical evolution 
one would expect slow prograde
satellites, which experience the largest dynamical friction, to be underrepresented. 
In actuality, all of these effects (and others) compete, and so detailed numerical simulations 
in a cosmological framework are necessary to make quantitative predictions.

Simulations attempting the necessary level of detail are overwhelmingly expensive
because of the required resolution over cosmological volumes.
This difficulty becomes particularly acute when one realizes that 
defining the observational sample
requires careful selection from large redshift  catalogues
\citep[see][hereafter B07]{B07} to ensure that one has found 
the dynamically isolated systems, and hence that similar volumes must 
be simulated to identify a truly corresponding sample in simulations. 
It is insufficient simply to identify random disc galaxies in simulations 
because the criteria for defining the observational sample is far 
more restrictive.  
Until recently \citep[][hereafter L07]{Okamoto, Governato07, L07}, 
$\Lambda$CDM galaxy formation simulations had been unable to create 
satellite populations of resolved and realistic spiral galaxies with properties 
similar to those of the Local Group.
Because the simulations are currently unable to provide
reliable guidance on dynamical signatures of galaxy assembly and 
formation that might be present in the most isolated systems residing in 
large-scale cosmological volumes, 
we search for any signatures in observational samples.  

One such signature that has received significant attention 
is the angular anisotropy in the distribution 
of satellites around isolated disc 
galaxies \citep[e.g.][]{holmberg,ZSFWb,SL04, Brainerd05}, 
often referred to as the Holmberg effect, even though the
results regard the satellite distribution over a significantly different, and possibly physically distinct,
range of projected radii than that explored in the original study.
This anisotropy, if confirmed, likely has its roots in the anisotropic infall of satellites 
along filaments \citep[e.g.][]{Aubert04, Knebe04}, but simulations have so
far failed to fully explain it primarily because they have difficulties in realistically
forming the central galaxies and satellite populations in a cosmological context.
The possible detection of satellite anisotropy to large 
radii \citep[out to $\sim 500\ h^{-1}_{75}$ 
kpc;][]{ZSFWb}, dynamical studies \citep{Prada06}, and weak lensing studies 
\citep{Guzik02} all suggest that the haloes of these galaxies
are quite extended, that there is a connection between the
central galaxy and the outer halo that extends even beyond the virial
radius,
and that these satellites may provide additional clues linking
the inner and outer galaxy.

Our goal is modest: we search for additional such signatures connecting 
the inner, observable, disc galaxy and the outer, dark halo by exploring the 
distribution of satellite angular velocities relative to the disc.
Most studies of satellite dynamics only measure the difference 
in recessional velocities, $\Delta v_{rec}$, between primary and satellite.
ZSFW, in addition,  measured the rotation sense of the primary
and were therefore able to measure, in projected space, whether a 
satellite was on a prograde or retrograde orbit. With
their limited sample size, they focused on measuring the prograde fraction, which
they found to be slightly above 0.5. Since ZSFW, both the angular anisotropy 
and rotation results have been revisited, with conflicting results 
\citep{SL04, Brainerd05, Azzaro06, WK06}.

B07 demonstrate that anisotropy measurements can be strongly affected
by the sample selection.  It is then also 
likely that sample selection is critical in a measurement of the
distribution of satellite orbits. Therefore, we focus on 
the two existing samples that B07 demonstrate, based on comparison
to mock galaxy  catalogues, truly select dynamically
isolated systems--the original ZSFW sample and the SDSS sample
constructed by B07.  By combining the results from the ZSFW study with
new rotation measurements of primaries drawn from B07, 
we increase the sample of satellites with measured orbital direction by 86, 
nearly doubling the ZSFW 
sample size (91 ZSFW satellites satisfy the selection criteria used here).
The sample selection is described in \S2, our 
observations and data reduction in \S3, results, 
in particular our findings regarding the dynamical connections
between the outer halo and inner galaxy, in \S4, a brief discussion in \S5,  and 
our summary and conclusions in \S6.

\section[]{Sample Selection}

The key ingredient in any dynamical analysis of a sample of satellite galaxies is the 
purity of the satellite sample. Purity requires a sample of isolated primary galaxies
and minimal contamination of the satellite population by interlopers along the line of sight. 
The value of large surveys like 2dFGRS and SDSS lies
primarily in the ability to accurately identify such isolated systems, 
rather than in the absolute increase in size of the satellite sample. In other 
words, the uncertainties inherent in certain previous results, such as 
the Holmberg effect \citep[e.g.][]{ZSFWb, SL04, Brainerd05}, 
are not principally statistical --- even the early work \citep{ZSFWb} 
detected the asymmetry with $> 99\%$ confidence at radii $> 250$ kpc.  
B07 demonstrated that conflicting results among studies \citep[][etc.]{SL04, Brainerd05} 
were caused by differences in the sample selection, not small number statistics.

We use a sample of nearby isolated galaxies selected from 
SDSS-DR4 (Adelman-McCarthy et al., 2006) by B07 with
3600 \kms $< v_{rec} < 38000$ \kms (no cut on depth beyond B07 
was made). The criteria B07 used to define 
this sample were refined using mock  catalogues 
generated from cosmological $N$-body simulations.  
B07 conclude that many previous 
studies are heavily contaminated by groups, and that the isolation 
criteria must be stringent in order to select truly isolated systems.  
Their isolation criteria are 
tailored to (1) minimize the number of `interlopers', or satellite-primary pairs that 
do not represent physical satellites of the primary galaxy, (2) minimize the number of 
primaries that are not isolated and so do not dominate the dynamics of their 
environment, and (3) maximize the sample size.  

The sample from which we selected targets for follow-up
was extracted from the SDSS-DR4 while the selection criteria 
of B07 were still being refined, and therefore there are slight 
differences between our sample and that of B07.  
The main differences are (1) our sample 
contains no constraint on proximity to the spectroscopic survey edge 
(although the more important constraint on proximity to the photometric 
survey edge is included), and (2) the issue of spectroscopic 
incompleteness (after checking NED\footnote{http://nedwww.ipac.caltech.edu/} for 
spectra of any potential violators not already in SDSS-DR4)
is dealt with differently.

The question is whether to exclude potential primaries on the basis of projected neighbors
that are sufficiently luminous to violate the isolation criteria, if they indeed
lie sufficiently close in redshift to the primary.
We, as is done in B07, refer to the
number of such galaxies around one of our primaries as $N_{\mathrm{viol}}$. 
ZSFW adopted a goal of $N_{\mathrm{viol}}=0$, although given their visual
magnitude classification and lack of supporting materials at the time it is quite possible
that some of their primaries do not satisfy this criteria. B07 also adopt  $N_{\mathrm{viol}}=0$,
but in an effort to minimize the number of systems rejected they use the photometric
redshift information provided by SDSS to reject only systems with 
violators that have a photometric redshift consistent with that of the primary. Given the
errors on photometric redshifts, this is still quite a conservative criteria because many of 
the potential interlopers will turn out to be at redshifts sufficiently distinct from that of the
primary. We were initially much more liberal, not having the photometric redshifts
when our sample was chosen. The critieria for our sample is 
 $N_{\mathrm{viol}} \le 4$. Now that photometric redshifts are available, we 
revisit our sample.
Of the 78 Bok-observed satellites in our sample,
17 have potential violators whose photometric redshift is consistent with that
of the primary. Eleven of the 12 primaries from which these 17 satellites are drawn 
are found to have only one such violator and one has 
two such violators, and so, with one exception, our sample satisfied $N_{\mathrm{viol}}\le 1$.
Our results are qualitatively unchanged when
these 17 satellites are excluded.

Lastly, we also require that
the maximum number of satellites allowed around any given 
primary, $N_{\mathrm{satmax}}$, is four.  This criterion
is set to ensure that the primary galaxy dominates the 
satellite system and  to eliminate systems that are clearly groups rather 
than isolated galaxies. 
ZSFW do not use such a criterion, and as a result one 
system in their final sample (NCG 1961) has 5 associated satellites.  
See $\S$~4.1
for a description of the (minimal) effects of this system on our results.

For completeness, we correct heliocentric $v_{rec}$ values for 
Galactic rotation and Virgocentric infall as described in \cite{ZSFW93},
although these corrections are negligible over these volumes and affect only
our determination of the distance to the system. 
Angular diameter distances and distance moduli are calculated assuming 
$\Omega_m$ = 0.3, $\Omega_{\Lambda}  = 0.7$ and $H_0 = 70$ \kms Mpc$^{-1}$.  

We work on a subset of the B07 sample that satisfies
three additional criteria. First, we visually select 
a strictly disc-like ($\sim$ S0 type or later) 
sample requiring the galaxies to appear to be inclined discs.  
This criterion is applied so that there is a well defined disc major axis and
the inclination can be unambiguously measured.
We then select primaries with
inclinations larger than 25\dg (where 0\dg is face-on) 
to increase our chances of measuring the disc rotation sense.
Finally, we remove any obviously disturbed systems 
(e.g.\ clear mergers or warps) 
because such systems, although perhaps satisfying the broader 
isolation criteria, are 
unrepresentative of the typical isolated galaxy in our sample and likely 
have ambiguous disc/halo alignments.  In all, these additional criteria 
yield a base sample of 249 primaries that we draw 
from during the observations (see $\S$~3).

We will combine the results obtained from this sample to the results presented by
ZSFW, which B07 confirm to have selection criteria resulting in a sample of
similarly isolated primary galaxies.

\section[]{Observations, Data Reduction \& Analysis}

The SDSS catalogues provide positions, magnitudes and radial velocities
for the primary and associated satellite galaxies in the B07 sample. 
To measure whether the satellite is on a prograde or retrograde orbit 
we need to measure the sense of the rotation of the primary galaxy on the sky.
We use the Boller \& Chivens (B\&C) Spectrograph 
on the Steward Observatory 90-inch (2.3 m) Bok Telescope (Kitt Peak, AZ) and 
observe 126 primary galaxies with 151 associated satellites 
($\sim 1.2$ satellites per primary) from the 249 primaries available 
in our base sample (see $\S$~2).
We use  the 600 g/mm grating blazed 
at 4458 \AA\ and centered at $\lambda \sim 5000$ \AA\ 
to observe a spectral range that includes  the Ca\,{\sevensize II} H \& K 
absorption lines and the H$\beta$ and [O\,{\sevensize III}] emission lines.
This setup provides a spectral scale of 1.9 \AA\ pixel$^{-1}$ 
and a spatial scale of $0.33^{\prime\prime}$ pixel$^{-1}$.  We use the 
$2.5^{\prime\prime}$ wide slit and 
achieve a typical spectral resolution of 
$\sim 6$ \AA.  Exposure times range between 1200 -- 1500 seconds. 
We obtained the data from three separate 
bright runs between October 2005 and April 2006.
Because we simply aim to determine which side of the galaxy is rotating 
toward or away from us, we do not require high precision or absolute
velocity calibrations.

To observe a galaxy rotation curve, one would normally align the spectrograph
slit with the major axis. Given our cruder requirements, 
our goal of maximizing the number
of observed targets, and the inefficient way in which the slit is rotated in
this spectrograph, we decided to observe all of our targets with a limited set of
slit angles. We chose to deviate by not more than 22.5\dg from the primary's major
axis and therefore used one of four slit 
angles 45\dg, 90\dg, 135\dg and 180\dg to observe every galaxy.
Because the entire B\&C Spectrograph rotates with the 
slit, we take calibration frames at each slit angle
to account for instrument flexure.

We reduce the images in the standard manner, including bias subtraction
and flat fielding using standard IRAF routines\footnote{IRAF is distributed 
by the National Optical Astronomy Observatories, which are operated by 
the Association of Universities for Research in Astronomy, Inc., under 
cooperative agreement with the National Science Foundation.}.
We rectify the images using a calibrated HeNeAr 
exposure and then subtract the background along columns. 
The center of the galaxy spectrum along the spatial
direction is identified by the APALL task and an aperture that is between
1 and 3 pixels wide is extracted. We generally adopt the 3 pixel aperture to
increase the signal-to-noise, but occasionally test whether smaller apertures result in
a more statistically significant final result. 
The extracted spectrum serves as the template for a 
cross-correlation analysis of the other rows. Row by row, we
cross-correlate the spectrum and the central template using XCSAO.  We accept
results only for rows with significant R values ($R > 4$) and then 
fit a line to the relative radial velocity as a function of row number.
A significantly non-zero slope provides a measure of the sense of 
rotation along the slit.

For quality assurance every derived rotation curve and 2-D spectrum 
is visually inspected.  To assist with the latter, we fit and subtract 
a low order continuum row-by-row.  This step increases the visibility of the
absorption or emission lines.  
If the correlation analysis appears to be strongly affected by an
artifact in the data (e.g.\ residual cosmic ray signal, bad pixels or columns), 
we reject this primary from the sample ($< 5\%$ of all cases).
We reach a final conclusion
regarding the rotation sense of the primary galaxy using both the cross-correlation
analysis and our visual inspection of the spectrum.  
We emphasize that there is no simple relationship between slit 
position or detected primary rotation sense and the location and
velocity of a satellite.  Our interactive analysis therefore can not incorporate any 
bias in the determination of the satellite orbit.

\begin{figure}
\vspace{0.8cm}
\hspace{-0.1cm}
\centerline{
\includegraphics[width=3.25in]{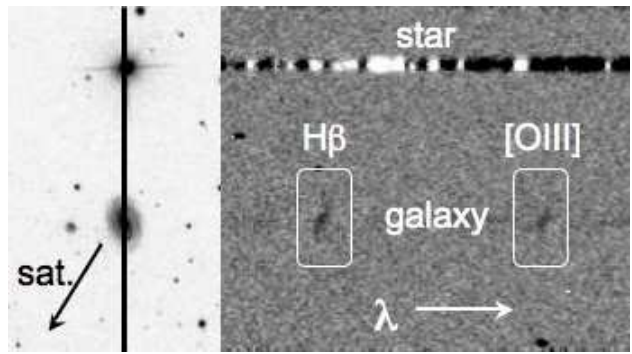}}
\caption{Combining the SDSS data and Bok observations.  On the left 
we show our slit/primary galaxy alignment (SDSS negative image), 
with an arrow indicating the direction of 
the associated satellite.  The right panel shows a cut 
of our Bok B\&C sky-subtracted spectrum, 
with the contaminant stellar spectra (note the star in the slit 
in the SDSS image to confirm the orientation; the stellar continuum appears 
unusual here due to the continuum subtraction), and 
the H$\beta$ and [O\,{\sevensize III}] emission lines of the galaxy below 
providing the primary's sense of rotation.  See the text for more details.
\label{slit_spec_fig}}
\end{figure}

We then combine the sense of primary rotation 
along the slit with the orientation of the slit on the sky, the position of the
satellite on the sky, and $\Delta v_{rec}$ to determine whether the
satellite is on a projected prograde or retrograde orbit. 
In Figure~\ref{slit_spec_fig}, we illustrate this process with 
an example merging the SDSS and Bok observations.  On the left 
we show our slit orientation overlayed on an SDSS negative 
image of the primary galaxy.
An arrow indicates the direction toward the associated satellite.  
The radial velocity difference between the satellite and this primary is 
$\sim -320$ \kms, i.e.\ the satellite is approaching us with respect to the 
primary.  In the right panel, we show a portion 
of our Bok B\&C sky-subtracted spectrum (longer wavelengths to the right), 
with a contaminating stellar spectra (having an unusual appearance due to the 
continuum subtraction) visible in the upper portion
and the H$\beta$ and [O\,{\sevensize III}] emission lines of the galaxy providing the primary's 
rotation sense visible in the lower portion.  
This particular example serendipitously includes the star which helps 
confirm our calculation of the slit orientation on the sky.
In this case, the spectrum of the lower side of the galaxy 
(that closest to the approaching satellite, or that toward the bottom of the 
CCD image) is blueshifted, indicating that the satellite is on a 
projected prograde $v_{r} \sim +320$ \kms orbit. Our adopted convention
is to use positive (negative) velocities,  $v_r$, to denote prograde (retrograde) orbits.

\begin{figure}
\hspace{-0.45cm}
\centerline{
\includegraphics[angle=90,width=3.5in]{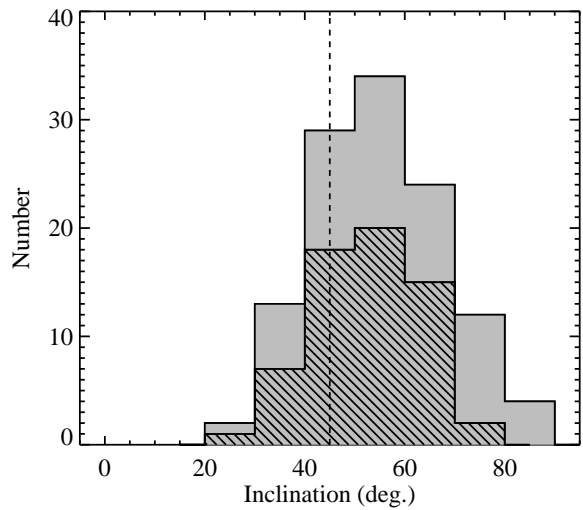}}
\caption{Distribution of inclination angles of the primary galaxies used 
in the analysis.  The full sample is shaded and those observed specifically for this
study are in hatched linestyle.
\label{inc_hists_base}}
\end{figure}

Of the 126 targeted primaries, we successfully measure 
the rotation sense of 63 primaries with 78 associated 
satellites. In Table 1 we present those primaries for which we were able to
measure a rotation sense and the information about their satellite systems.
This is the Bok-observed sample that we use in our analysis.
For completeness, in Table 2 we present the 63 primaries for which we were 
unable to measure a rotation sense. The majority of the failures 
appear to be weak-lined sources where our signal-to-noise was insufficient.
However, a significant fraction are also low-inclination ($inc < 50$\dg)
systems for which measuring a rotation sense is more difficult.
In Figure~\ref{inc_hists_base} we show the distribution of primary inclinations for the 
118 primaries we use in our analysis (shaded), which includes the 63 Bok-observed 
primaries (hatched linestyle) and the 55 ZSFW primaries.
Our sample is therefore 
not in any sense random in orientation and 
simulations that attempt to reproduce these observations should begin with
a sample that matches our distribution of primary inclinations.

\section[]{Results}

\subsection{Deviations from Gaussianity}

\begin{figure}
\hspace{-0.45cm}
\centerline{
\includegraphics[angle=90,width=3.5in]{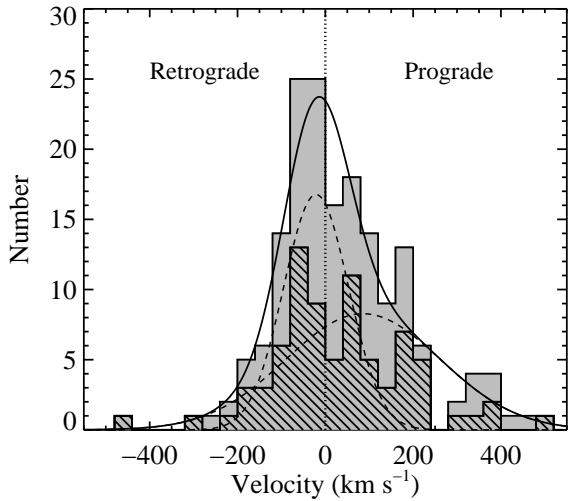}}
\caption{Radial rotation velocities in 40 \kms bins.  
The new Bok sample is shown in hatched linestyle and the overall (Bok + ZSFW) 
sample is shaded.  Positive values depict satellites rotating 
in the same sense as the disc, while negative values indicate counter-rotating 
satellites.  These data are inconsistent with a normal distribution at the 
$>99.9$ \% CL.  A double-Gaussian model provides an acceptable fit to 
the distribution (solid linestyle, with the individual Gaussian components in 
dashed linestyle). 
\vspace{0.05cm}
\newline Note: these fit parameters were derived from 
the {\it unbinned} $v_r$ distribution using a maximum likelihood approach.  
They are not a fit to the binned distribution shown here.
\label{V_hist_DISK_DG_mpfit}}
\end{figure}

In Figure~\ref{V_hist_DISK_DG_mpfit} we show 
the projected rotation velocities for satellites in the Bok-observed sample 
(hatched linestyle) and the combined Bok + ZSFW sample (shaded).
There are three general features that are present in both the Bok and
ZSFW samples: 1) a deviation from Gaussianity, 2)  an asymmetric tail 
toward large prograde velocities, and 3) a slightly off-center peak at low 
retrograde velocities.
We confirm the visual impression of non-Gaussianity using the 
Bera-Jarque test for normality \citep{BJtest}, which uses the 
unbinned  $v_r$ distribution, at the $> 99.9$\% confidence level.  

We explore the asymmetry further by testing the statistical significance of the asymmetric tails.
We calculate the probability, for the combined sample, 
of finding P or more satellites with $v_r > $ X \kms and N or less satellites
at $v_r <$ $-$X \kms using Binomial probabilities and adopting 
an equal chance for positive and negative velocities. Of course, 
this is an {\sl a posteriori} test if we choose a particular
value of X and hence we do the calculation for all values 
$0  < {\rm X} < 300$ \kms (for X $>$ 306 \kms
we have ten or fewer satellites). For values of X $> 50$ \kms we find that
the probability of randomly generating 
the observed asymmetry toward prograde orbits is $< 5$\%, while for 
X $> 60$ \kms the probability is always $<$ 2 \%. We conclude that the asymmetry is
statistically significant under the assumptions that prograde and retrograde orbits are
equally likely and that each satellite is an independent measurement.

However, the assumption that each satellite is an
independent dynamical tracer is questionable.  Two of the 
systems among the largest prograde velocities
are from a single primary, ZSFW's NGC 1961, which
was identified by ZSFW as being notably different than the rest of the sample
because of its unusually large circular velocity (551 \kms width of the neutral 
hydrogen profile at 20\% peak intensity) and large number of satellites (five).
If we remove the NGC 1961 satellite system from the sample (it has one more satellite
than the cut imposed by B07 for their sample, $N_{\mathrm{satmax}}=4$), we calculate 
that for $55 < {\rm X} < 300$ \kms the probability of 
the asymmetry remains $<$ 5\%.  Interlopers, which are expected at the 
$\sim 5-10$\% level \citep{B07}, 
will not correlate with the primary's sense of rotation and therefore will not
produce $v_r$ asymmetries.
We therefore conclude that the asymmetries at large $v_r$ are statistically significant
and physical.

We return now to the peak at low retrograde velocities, which is evident in both 
the Bok and ZSFW samples.  In Figure~\ref{V_hist_DISK_DG_mpfit} we show 
the result of fitting a double-Gaussian model to the {\it unbinned}
$v_r$ distribution using a maximum likelihood approach.
We find that the distribution can be described as the sum
of a slightly retrograde component containing 
$\sim 45$\% of the sample peaked at $-21 \pm 22$ \kms 
with standard deviation $\sigma = 74 \pm 18$ \kms, and a 
broad prograde component containing 
$\sim 55$\% of the sample peaked at $86 \pm 30$ \kms 
with $\sigma = 176 \pm 15$ \kms.
These parameters agree to $\sim 1 \sigma$ with the values derived
from a double-Gaussian best-fit to the binned data that has a 
reduced $\chi^2 \sim 0.6$.
The maximum-likelihood fit parameter values derived from 
just the Bok- and ZSFW-only subsamples 
are consistent with those from the main sample.  

\begin{figure}
\hspace{-0.45cm}
\centerline{
\includegraphics[angle=90,width=3.5in]{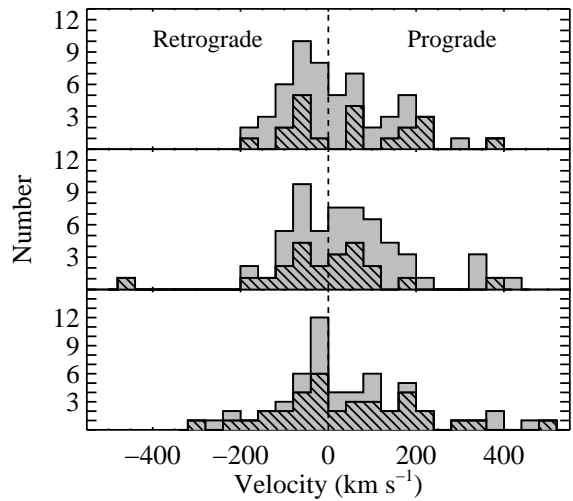}}
\caption{
Rotational velocity distributions in projected radii bins 
(from top) [0 -- 162], [162 -- 368], and [368 -- 750] kpc, each 
containing 56, 56, and 57 satellites, respectively.  The overall 
asymmetric distribution is apparent throughout the halo.
\label{V_hist_rad_all}}
\end{figure}

\begin{figure}
\hspace{-0.45cm}
\centerline{
\includegraphics[angle=90,width=3.5in]{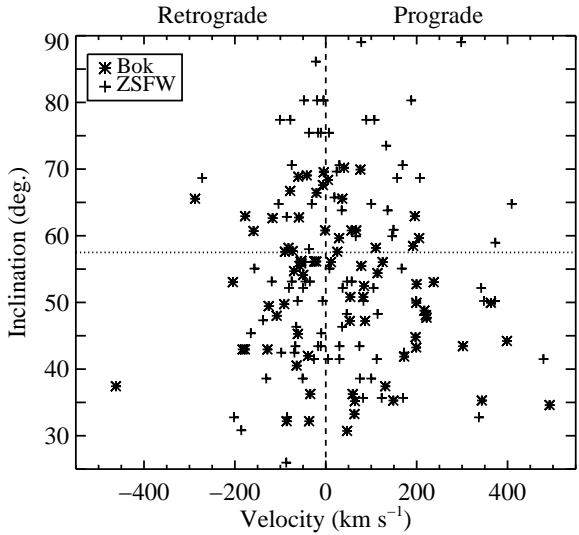}}
\caption{Rotational velocity vs.\ primary inclination angle.  
A dotted line is drawn at 57.5\dg (the midpoint of the 
inclination range considered) to guide the eye.
\label{vrot_inc_DISK_base}}
\end{figure}

\begin{figure}
\hspace{-0.45cm}
\centerline{
\includegraphics[angle=90,width=3.5in]{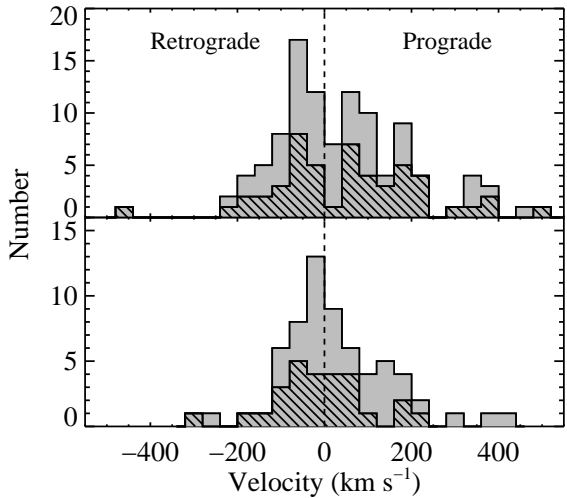}}
\caption{Rotational velocity distributions in inclination bins 
(from top) [25\dg -- 57.5\dg], [57.5\dg -- 90\dg].  As 
Figure~\ref{vrot_inc_DISK_base} suggests, the overall 
asymmetric distribution is apparent across the range of 
primary inclinations.
\label{V_hist_inc_all}}
\end{figure}

\begin{figure}
\hspace{-0.45cm}
\centerline{
\includegraphics[angle=90,width=3.5in]{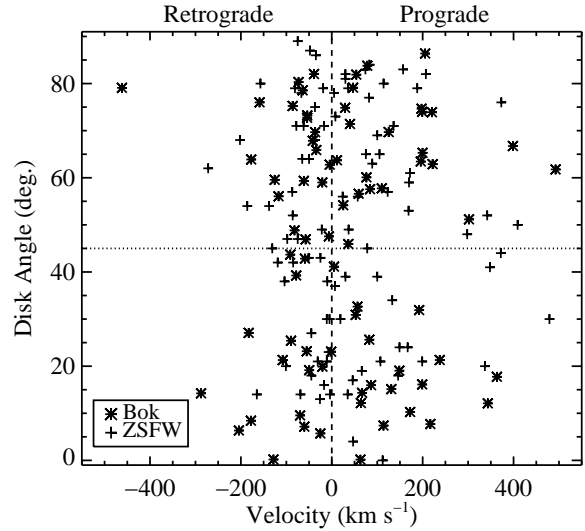}}
\caption{Rotational velocity vs.\ projected satellite disc angle (angle 
between the satellite and the nearest semi-major axis of the host primary).  
A dotted line is drawn at 45\dg to guide the eye.
\label{vrot_dang_DISK_base}}
\end{figure}

\begin{figure}
\hspace{-0.45cm}
\centerline{
\includegraphics[angle=90,width=3.5in]{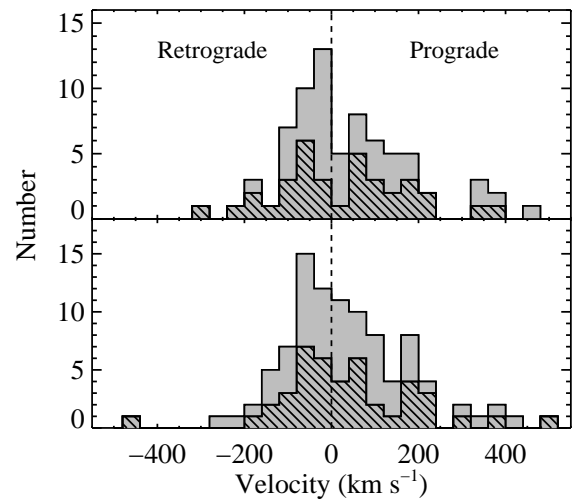}}
\caption{Rotational velocity distributions binned by satellite disc angles 
(from top) [0\dg -- 45\dg], [45\dg -- 90\dg].  As 
Figure~\ref{vrot_dang_DISK_base} suggests, the overall 
asymmetric distribution is apparent across the range of 
disc angles.
\label{V_hist_dang_all}}
\end{figure}

In Figure~\ref{V_hist_rad_all} 
we plot the $v_r$ distribution in three radial bins to illustrate that the 
asymmetric shape is found at all radii.
The maximum-likelihood double-Gaussian fit parameters in each radial
bin are roughly consistent (between $1 - 2 \sigma$) with those from the main sample.
We also look for any trends with primary inclination and disc angle 
(the latter being the angle between the satellite position on the sky 
and the nearest semi-major axis of the host primary disc) 
and find these slow retrograde satellites scattered throughout 
(see Figures~\ref{vrot_inc_DISK_base} -- \ref{V_hist_dang_all}).

\begin{figure}
\hspace{-0.45cm}
\centerline{
\includegraphics[angle=90,width=3.5in]{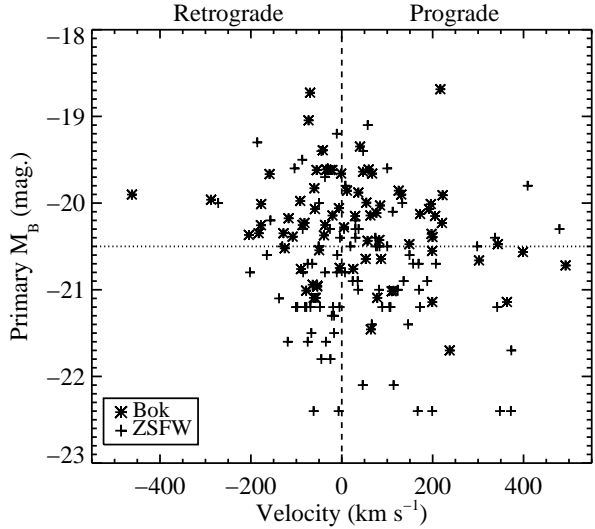}}
\caption{Rotational velocity vs.\ primary absolute magnitude.  The median 
primary M$_B$, $-20.5$ mag., is traced by the dotted line.
\label{vrot_pMB_DISK_base}}
\end{figure}

\begin{figure}
\hspace{-0.45cm}
\centerline{
\includegraphics[angle=90,width=3.5in]{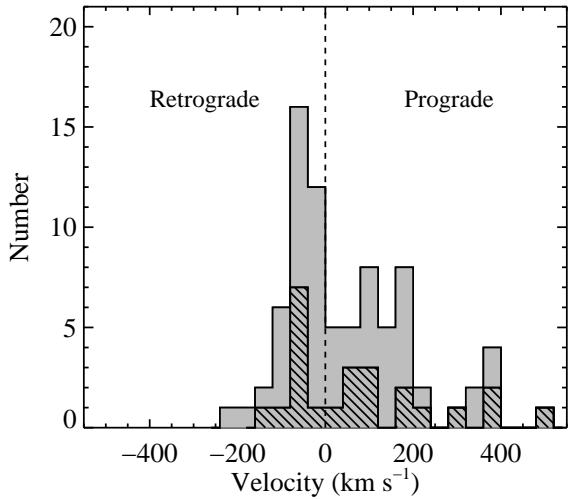}}
\caption{Rotational velocity distribution for satellites around the `bright' primary subsample.  
The narrow retrograde and broad prograde components are clearly distinguished here.
\label{v_hist_bright}}
\end{figure}

\begin{figure}
\hspace{-0.45cm}
\centerline{
\includegraphics[angle=90,width=3.5in]{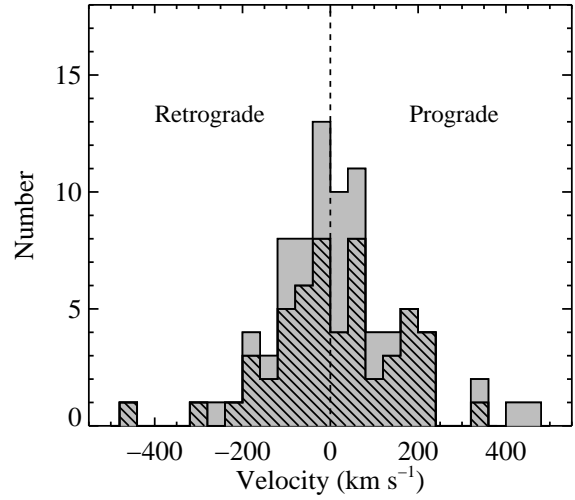}}
\caption{Rotational velocity distribution for satellites around 
the `faint' primary subsample.  The unbinned data here remain 
inconsistent with a single Gaussian at the $>99$\% CL, but there
is little evidence for the two components.
\label{v_hist_faint}}
\end{figure}

The only suggestion of a dependence we find 
between the satellite velocity distribution and
primary galaxy property is that with primary absolute magnitude
(see Figure~\ref{vrot_pMB_DISK_base}), in which the asymmetry toward
large prograde velocities becomes more pronounced for
brighter primary galaxies (we find no dependence on 
satellite absolute magnitude or primary-satellite magnitude 
difference). Dividing the sample 
into `bright' and `faint' subsamples on
either side of the median primary M$_B$, $-20.5$ mag, illustrates that it is
the bright subsample that more distinctly follows the pattern of the overall
two-component distribution (Figure~\ref{v_hist_bright}). 
The satellite velocity distribution for the faint primary subsample
(Figure~\ref{v_hist_faint}) is closer to that of a single Gaussian, 
although using the statistical test for normality on 
the unbinned values shows that even this distribution is inconsistent with a
single Gaussian at the $> 99$\% CL. These results may reflect a true
physical difference among satellite orbits of bright and faint primary galaxies, or 
the ease with which one can distinctly separate the
prograde component for the brighter primaries, which will be centered at
larger mean velocity in the more luminous systems \citep[see e.g.][]{Guzik02}.

\subsection{The Normalised Distribution}
When stacking objects it is always difficult to know if one can simply combine
the sample without accounting for the differences among individual objects.
In our case, we have combined satellites of primaries with different luminosities,
and hence rotation speeds. One natural alternative to our straight-forward stacking 
is to rescale the satellite-primary
velocity differences by primary rotation speed before stacking.
We estimate the primary's circular velocity using the Tully-Fisher (TF) 
relation of \cite{Pizagno07}, log($V_{80}$) $= -0.136 \times 
({\rm M_{g}} + 20.607)+2.209$, where $V_{80}$ is approximately 
the circular velocity, $v_{circ}$, and we have replaced ${\rm M_g}$ 
with our ${\rm M_B}$ values. 

\begin{figure}
\hspace{-0.45cm}
\centerline{
\includegraphics[angle=90,width=3.5in]{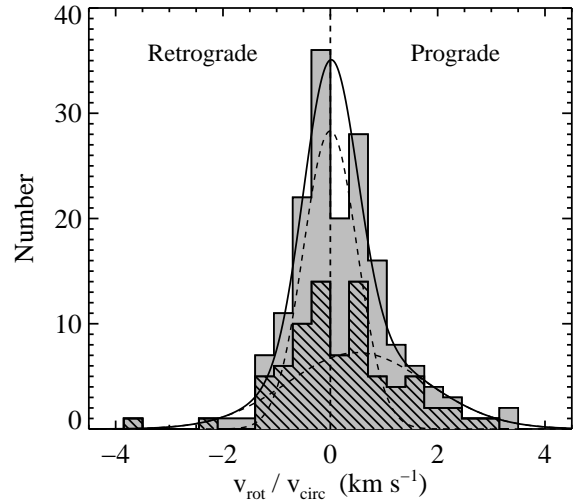}}
\caption{
Projected orbital velocity distribution, normalised using the Tully-Fisher 
relation of Pizagno et al.\ (2007), in 0.35 $v_{rot}/v_{circ}$ bins.  The 
double-Gaussian fit was found using a maximum likelihood approach to 
the {\it unbinned} $v_{rot}/v_{circ}$ distribution and is not a 
fit to the binned distribution shown here.
\label{v_hist_TFnorm}}
\end{figure}

The normalised satellite velocity distribution, shown in 
Figure~\ref{v_hist_TFnorm}, is still asymmetric 
and remains inconsistent with a normal distribution at the $> 99.9$\% CL.  
For completeness, we present the double-Gaussian best-fit parameters 
derived from the {\it unbinned} distribution using a maximum likelihood approach:  we 
find the left-most peak centered at $0.0 \pm 0.2$ with standard deviation 
$\sigma = 0.6 \pm 0.1$, and the broad prograde peak 
at $0.5 \pm 0.3$ with standard deviation $\sigma = 1.3 \pm 0.2$. These results
are consistent both qualitatively and quantitatively with the result from the 
straight-forward stacking analysis.

\subsection{Comparison with Previous Results}

There have been only two studies measuring the orbital sense 
of satellites \citep{ZSFWb, Azzaro06}, one of which we have incorporated
here, so there is not much comparison possible. However, 
one aspect to expand upon is the method in which the distribution
should be parametrized.  ZSFW presented the fraction of satellites on prograde 
orbits, which was also used by \cite{Azzaro06} and is $0.53 \pm 0.04$ 
in our sample.  Our new value is lower, but 
roughly consistent (between 1 and 2$\sigma$) 
with results from the previous studies.  

\cite{WK06} conducted high-resolution $N$-body simulations of dark matter haloes in a
standard $\Lambda$CDM cosmology and examined the velocity 
distribution and orientation of orbits of satellite galaxies.  Their general result 
was an excess of prograde satellites; on average they find that $59\%$ of 
satellites co-rotate with the host.  However, \cite{WK06} also find that the 
prograde fraction decreases to $53\%$ when observing the data 
along a random line of sight, as we (nearly) do in practice (any biased 
distribution of primary inclinations angles should still be accounted for, 
of course).  The decline of the prograde fraction is due to 
the technique observers use to determine the orbital sense of satellites 
and results in the possible misclassification of non-circular orbits as retrograde when 
in fact they are prograde (and vice versa).  
Thus, our measurement of a positive satellite prograde fraction 
(and those of any observers with a similar technique) 
may likely be an underestimate.

Alternatively, a useful value to compare with is the 
mean rotation velocity, which we find to be $37 \pm 3$ \kms 
($>12\sigma$ discrepant with zero). Note that this mean value 
must in fact be a lower limit to the coherent motion aligned with the disc 
because the primaries are not fully edge-on, the line-of-sight satellite 
velocities are not their full orbital velocities, and any interlopers are diluting 
the signal.

While both of these measures suggest the existence of
bulk mean rotation of the outer dark halo in the same sense as the disc, 
they both fail to capture the nature of the
deviations from the naively expected $v_r$ distribution.
In fact, had the fraction of satellites present in the narrow and prograde
peaks been slightly different, it is quite possible that neither the prograde
fraction or the mean velocity would provide statistically compelling evidence
against a purely Gaussian, non-rotating halo. The full $v_r$ distribution must
be measured and modeled.

\section[]{Discussion}

The presence of asymmetries in the $v_r$ distribution
confirms a connection between the inner, visible disc and the outer, dark halo. 
This connection reaches even beyond the galaxies' virial radii 
(indeed, Prada et al.\ 2006 also find that the haloes of simulated 
isolated galaxies may extend out to $\sim3$ virial radii, and 
\cite{WK06} find a bulk mean prograde rotation sense of subhaloes 
to five virial radii), further suggesting that large-scale structure plays a detectable 
role in determining both the disc orientation and 
the kinematics of the outer halo.  Large-scale structure determines 
the tidal field around galaxies, and both the gas in host discs and the orbits 
of the satellites may plausibly reflect torques from this large scale 
structure (such as from nearby clusters or filaments).  The connection is
potentially extremely valuable for constraining cosmological 
galaxy formation models because
it ties together the physical processes that form the central galaxy and 
the infall pattern of nearby larger-scale structures.  The complicated baryonic 
processes involved in forming the inner galaxy do not appear to erase
this connection.  

We briefly note here the work of \cite{Diemand04}, who identified 
dynamical signatures imprinted on the final velocity distribution of subhalos.
They found, in simulations of cluster-sized dark matter haloes, that substructure
has a flat-topped, non-Maxwellian velocity distribution. 
On this  scale,  which is not necessarily related to our systems, \cite{Diemand04} find a 
{\it lack} of slow subhaloes, which they in part attribute to the tidal disruptions of slow substructure 
in the dense, young cluster environment.  Analogous investigations of isolated, galaxy-sized
haloes may provide insights into our findings.

The very recent work by \cite{Sales07} examines satellite 
dynamics in isolated galaxy halos drawn 
from the Millennium Simulation, and hints at the way forward in 
terms of modeling the results presented 
here.  While the cosmological volume covered by the Millennium 
Simulation is adequate, the semi-analytic population and evolution of 
the host halos does not provide an `observer' with the orbital velocity 
distribution of the satellites.  Nevertheless, it is important to note 
that \cite{Sales07} find an asymmetric satellite {\sl radial} velocity 
distribution with a distinct double-Gaussian shape in the outer halo regions 
(outside the virial radius) and a sharp peak at low 
negative (not retrograde) velocities.  The authors attribute this 
component to infalling satellites, and the broader single-Gaussian 
shape (which dominates the inner regions) to a relaxed satellite population in 
equilibrium.  While it is not clear if or why the sharp peak at low 
negative radial velocities would transform into the peak we see at slow 
retrograde velocities (when taking into account the 
primary orbital sense, which is not provided in the simulation), it is nevertheless very 
interesting to find a distinctly non-Gaussian distribution in the simulations. 

Given the current limitations of the simulations, 
we are left to contemplate the empirical findings without 
much of a predictive framework.  
The question is then whether the double-Gaussian we use to model 
the $v_r$ distribution
is simply  a convenient way to model a complex distribution or faithfully represents 
the presence of two physically distinct components.
We speculate that the broad, prograde component is the classic halo. Its
$v/\sigma$, 0.49, is similar to that of our Galaxy as traced by 
metal-poor globular clusters \citep[0.44; ][]{Zinn85}.  

The origin of the slightly retrograde peak, albeit  consistent with 
zero rotation, is more mysterious. Although
dynamical friction is stronger for prograde orbits, 
and should thereby leave an excess of retrograde orbits, the effect is expected to 
be weak outside of 50 kpc \citep{Quinn86}.
We find that the excess of slow retrograde orbits extends well beyond
this radius, and hence conclude that dynamical friction is not
principally responsible. Given their low $v_r$, these satellites are perhaps a 
population on radially biased orbits and we may then be seeing the remnants
of the radial population (Figure~\ref{V_hist_rad_all} does suggest 
that the retrograde peak is most pronounced at the largest projected radii).  
If indeed the mean rotation is not retrograde, but rather consistent with no
net rotation, then this component may be associated with the late infalling 
satellites identified in the simulations by \cite{Sales07}.

Under the assumption that the broad prograde component is a reliable 
indicator of the overall halo rotation, the fact that the velocity dispersion remains 
nearly unchanged when considering either the rotational or line-of-sight (LOS) 
velocity distributions indicates that the results presented here will have minimum impact on 
previous halo virial mass estimates from satellite studies \citep[e.g.][]{ZSFW93, ZW94}.  
In other words, the mean rotation is sufficiently smaller than the dispersion
that unknowingly folding in the rotation velocity to the velocity dispersion
does not give rise to significant errors. To be more quantitative, we find that
the error introduced into the measured velocity dispersion by the rotation 
is less than the 1$\sigma$ uncertainty in the velocity dispersion measurement. 

\begin{figure}
\hspace{-0.45cm}
\centerline{
\includegraphics[angle=90,width=3.5in]{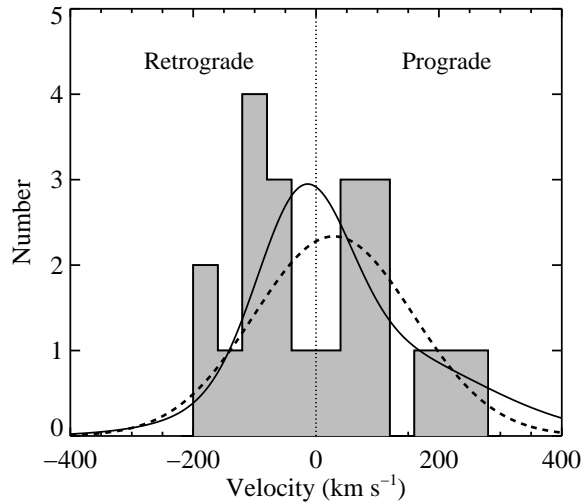}}
\caption{Projected rotational velocity distribution for satellites around M31 (see text for 
references).  The rescaled 
single- and double-Gaussian fits from the ensemble sample are overplotted 
in dashed and solid linestyle, repsectively.  The 
small sample size precludes any distinction between these two models.
\label{M31_sats}}
\end{figure}

Because of the ensemble stacking method used here, we do not discount 
the possibility that the observed bimodal $v_r$ distribution represents two separate 
{\it primary} galaxy populations, rather than two perhaps distinct satellite populations.  
While it is not obvious why some primaries would tend to 
form with more slow retrograde satellites, for example, seeing whether or not 
single galaxies with sizable satellite populations like the Milky Way or M31 exhibit similar 
double-peaked satellite $v_r$ distributions is a test of this interpretation.
A first attempt at this test, using data of 19 M31 satellites from 
\cite{McConnachie06} and the two 
recently discovered satellites And XII \& XIV with measured radial 
velocities \citep{Chapman07, Majewski07}, is presented in Figure~\ref{M31_sats}.  
Despite the irregular appearance of the distribution, a Bera-Jarque 
test for normality on the unbinned distribution does not reject the hypothesis
that these data are drawn from a Gaussian distribution. 
Rescaled and overplotted in Figure~\ref{M31_sats} are our 
best-fit single- and double-Gaussian models to our ensemble sample 
(in dashed and solid linestyle, repectively).  Neither model can be rejected with
$> 2 \sigma$ confidence, although the single Gaussian is the slightly better fit.
The current sample is still too small to resolve this question.

Finally, we also caution that the dearth of slow 
prograde satellites may be an artifact of the
selection of primaries that exhibit no signs of disturbance.  Recall that we exclude 
any obviously disturbed or warped systems from our final sample.  
If slow prograde satellites cause more morphological damage to their primaries, then 
perhaps these disturbed systems preferentially host a majority of (observable) 
slow prograde satellites.  However, the asymmetric $v_r$ 
distribution is apparent throughout 
the halo to the largest radii, where the effects of satellites on the 
primary are negligible.

\section[]{Summary and Conclusions}

We use a sample of isolated disc galaxies drawn from the SDSS-DR4 \citep[see][for details on selection criteria ]{B07} and 
Zartisky et al.\ (1997) to examine the distribution of satellite orbits
and uncover links between the dynamics and accretion history
of the outer, dark halo and the inner, visible disc.
We measure the rotation sense of 63 primary galaxies with 
78 associated satellites using the Steward Observatory 2.3 m Bok telescope 
and a long-slit spectrograph.
Combining these data with existing radial velocity measurements, 
we construct the projected prograde/retrograde orbital distribution of the ensemble 
satellite system around an isolated,  primary galaxy.   When combined with the
ZSFW study (91 satellites) the total sample has 169 satellites.  
We find a mean net bulk rotation 
of the satellites in the same sense as the primary, co-rotating with the host at 
$37 \pm 3$ km/s ($> 12\sigma$ from zero).  
We caution that no corrections for 
interlopers have been made, but that the sample selection was tailored 
to minimize the effects of interlopers.
We also find a prograde fraction of $0.53 \pm 0.04$ and caution that this is likely an 
underestimate of the true value \citep[see][]{WK06}.

The most interesting and constraining aspect of the data, however, 
lies rather in the {\it shape} of the projected rotational velocity distribution.  
We find that the data is distinctly non-Gaussian at the $>99.9 \%$ CL, 
and that a double-Gaussian model with a slightly retrograde component 
peaked at $-21 \pm 22$ \kms and with standard deviation 
$\sigma = 74 \pm 18$ \kms, together with a broader prograde 
component centered at $86 \pm 30$ \kms and with $\sigma = 176 \pm 15$ \kms 
provides an acceptable fit to the overall projected rotational velocity distribution.
The asymmetric shape is present throughout the sample, 
regardless of primary inclination, projected disc angle, or radial separation, 
yet becomes more distinct when considering satellites of the `bright' 
(M$_B < -20.5$ mag.) primaries.  The nature of these results is unchanged when
we scale the orbital velocities by the primary's circular velocity.
We hypothesize that the 
two components reflect physically distinct satellite families, perhaps as
suggested in a different context by \cite{Sales07}, but also discuss
whether they may reflect differences among the primary galaxies.

Given the kinematic asymmetry identified here,
it is evident that there remain significant 
underappreciated clues in the properties of satellites regarding the formation
of extremely isolated galaxies.

\section*{Acknowledgments}
We thank the SDSS team for their impressive survey 
and Chris Power and Peder Norberg 
for helping create the mock  catalogues used to define the isolated primary sample.  
SHF thanks Stephanie Cortes for useful discussions.
SHF, DZ, and YJK were partially supported by NSF grant AST-0307482 and 
NASA grant LTSA NNG05GE82G. DZ gratefully acknowledges 
financial support during his sabbatical from the 
John Simon Guggenheim foundation, KITP through
its support form the National Science Foundation grant PHY99-07949, and
the NYU Physics Department and Center for Cosmology and Particle Physics.  
JB thanks the Australian Research Council for financial support, and 
JET gratefully acknowledges support from NSERC Canada.

Funding for the Sloan Digital Sky Survey (SDSS) has been provided by the Alfred P. Sloan Foundation, the Participating Institutions, the National Aeronautics and Space Administration, the National Science Foundation, the U.S. Department of Energy, the Japanese Monbukagakusho, and the Max Planck Society. The SDSS Web site is http://www.sdss.org/.
The SDSS is managed by the Astrophysical Research Consortium (ARC) for the Participating Institutions. The Participating Institutions are The University of Chicago, Fermilab, the Institute for Advanced Study, the Japan Participation Group, The Johns Hopkins University, the Korean Scientist Group, Los Alamos National Laboratory, the Max-Planck-Institute for Astronomy (MPIA), the Max-Planck-Institute for Astrophysics (MPA), New Mexico State University, University of Pittsburgh, University of Portsmouth, Princeton University, the United States Naval Observatory, and the University of Washington.

\begin{table*}
 \centering	
 \begin{minipage}{119mm}
  \caption{Bok-Observed Primary/Satellite Sample. 
  RA \& Dec listed in degrees; absolute magnitudes are on the 
Vega system and determined using {\it{sdss2bessel}} from 
KCORRECT (Blanton et al.\ 2003) v4\_1\_4 with SDSS redshifts and dereddened 
apparent magnitudes; inc$_{pri}$ are the
inclinations of the primaries in degrees; $r_p$ are the projected radii
of the satellites from their host primary in kpc; $\theta$ are the projected 
disc angles of the satellites with respect to their host primary's 
nearest semi-major axis, in degrees; and 
$v_r$ are the final projected rotational velocities determined 
for the satellites, in \kms
(positive values indicate projected prograde rotation).}
  \begin{tabular}{@{}rrrrcccrrr@{}}
  \hline
   RA$_{pri}$ & Dec$_{pri}$ & RA$_{sat}$ & Dec$_{sat}$ & M$_{B(pri)}$ & M$_{B(sat)}$ & inc$_{pri}$ & $r_p$ & $\theta$ & $v_r$\\
 \hline
2.805 & 14.282 & 2.833 & 14.339 & -21.0 & -18.5 & 58 & 343 & 58 & 110 \\
16.600 & -0.145 & 16.599 & -0.170 & -20.3 & -17.4 & 43 & 84 & 8 & -178 \\
18.703 & -0.496 & 18.681 & -0.549 & -20.4 & -17.3 & 44 & 138 & 75 & 198 \\
23.230 & 14.969 & 23.159 & 14.938 & -20.6 & -18.5 & 44 & 367 & 67 & 399 \\
23.281 & 14.761 & 23.100 & 14.727 & -20.2 & -17.7 & 48 & 700 & 74 & 220 \\
27.783 & -9.792 & 27.753 & -9.761 & -21.1 & -17.9 & 45 & 131 & 59 & -62 \\
116.014 & 33.077 & 116.202 & 33.039 & -20.7 & -18.6 & 43 & 630 & 51 & 302 \\
116.110 & 29.269 & 116.174 & 29.316 & -19.9 & -17.2 & 56 & 206 & 64 & 11 \\
116.110 & 29.269 & 116.312 & 29.250 & -19.9 & -17.7 & 56 & 499 & 70 & 125 \\
117.801 & 22.423 & 117.773 & 22.367 & -20.2 & -17.7 & 58 & 199 & 49 & -82 \\
119.194 & 31.929 & 119.223 & 31.925 & -20.1 & -18.1 & 63 & 96 & 43 & -59 \\
120.764 & 27.744 & 120.761 & 27.584 & -20.1 & -17.6 & 59 & 534 & 32 & 192 \\
121.249 & 34.016 & 121.371 & 33.984 & -20.4 & -18.1 & 53 & 458 & 6 & -204 \\
123.003 & 39.013 & 123.060 & 39.049 & -20.0 & -17.7 & 51 & 167 & 82 & 54 \\
123.802 & 41.215 & 123.811 & 41.098 & -21.1 & -18.9 & 50 & 502 & 74 & 199 \\
123.802 & 41.215 & 123.766 & 41.194 & -21.1 & -18.2 & 50 & 147 & 18 & 363 \\
125.159 & 36.885 & 125.330 & 36.885 & -19.0 & -16.7 & 58 & 297 & 80 & -73 \\
127.217 & 42.522 & 127.252 & 42.520 & -20.1 & -18.0 & 41 & 101 & 10 & 172 \\
131.207 & 43.450 & 131.215 & 43.466 & -20.8 & -17.8 & 58 & 63 & 25 & -90 \\
131.207 & 43.450 & 131.415 & 43.429 & -20.8 & -18.1 & 58 & 578 & 54 & 25 \\
131.771 & 41.715 & 131.770 & 41.702 & -20.5 & -18.0 & 54 & 61 & 19 & -50 \\
132.014 & 37.663 & 132.251 & 37.574 & -20.4 & -18.0 & 42 & 710 & 82 & -39 \\
136.903 & 3.393 & 137.101 & 3.368 & -19.6 & -16.7 & 56 & 186 & 6 & -25 \\
136.903 & 3.393 & 137.213 & 3.448 & -19.6 & -17.7 & 56 & 292 & 23 & -55 \\
136.903 & 3.393 & 136.764 & 3.818 & -19.6 & -15.9 & 56 & 415 & 59 & -21 \\
137.973 & 37.404 & 137.889 & 37.366 & -21.7 & -18.9 & 53 & 530 & 21 & 237 \\
144.477 & 6.654 & 144.462 & 6.622 & -21.0 & -18.4 & 56 & 172 & 73 & -54 \\
144.516 & 42.974 & 144.572 & 42.973 & -20.4 & -18.0 & 61 & 135 & 33 & 57 \\
144.747 & 6.955 & 144.503 & 6.924 & -19.7 & -16.4 & 61 & 297 & 76 & -159 \\
144.894 & 3.158 & 144.999 & 2.996 & -20.1 & -18.0 & 68 & 633 & 48 & -6 \\
144.908 & 59.791 & 144.546 & 59.746 & -20.4 & -18.2 & 48 & 633 & 21 & -108 \\
147.734 & 62.186 & 148.175 & 62.281 & -19.4 & -16.3 & 69 & 404 & 68 & -42 \\
148.984 & 53.694 & 149.192 & 53.686 & -20.2 & -17.5 & 63 & 380 & 56 & -117 \\
151.007 & 38.677 & 150.898 & 38.621 & -20.1 & -17.5 & 70 & 343 & 60 & 76 \\
153.586 & 49.511 & 153.640 & 49.389 & -21.0 & -18.2 & 67 & 536 & 39 & -78 \\
154.085 & 4.822 & 153.949 & 4.750 & -21.0 & -17.6 & 56 & 501 & 73 & -54 \\
154.245 & 49.627 & 154.016 & 49.818 & -20.6 & -17.6 & 47 & 713 & 16 & 86 \\
154.245 & 49.627 & 154.197 & 49.630 & -20.6 & -18.0 & 47 & 91 & 31 & 53 \\
154.271 & 54.821 & 154.634 & 54.812 & -20.4 & -17.1 & 53 & 572 & 65 & 200 \\
154.971 & 55.436 & 154.689 & 55.494 & -19.6 & -16.8 & 36 & 397 & 57 & 59 \\
154.971 & 55.436 & 155.245 & 55.349 & -19.6 & -16.9 & 36 & 417 & 66 & -34 \\
156.430 & 39.646 & 156.397 & 39.674 & -20.1 & -18.2 & 33 & 64 & 0 & 63 \\
157.959 & 54.355 & 157.951 & 54.403 & -19.8 & -17.8 & 69 & 162 & 7 & -60 \\
158.333 & 58.076 & 158.074 & 58.230 & -20.0 & -18.2 & 52 & 663 & 58 & 84 \\
158.563 & 52.871 & 158.550 & 52.926 & -20.1 & -16.9 & 67 & 95 & 20 & -20 \\
160.065 & 65.492 & 160.343 & 65.395 & -19.9 & -17.8 & 66 & 367 & 46 & 36 \\
165.816 & 54.111 & 165.883 & 54.119 & -21.1 & -18.7 & 55 & 194 & 47 & -58 \\
165.816 & 54.111 & 165.759 & 54.074 & -21.1 & -18.4 & 55 & 242 & 84 & 78 \\
\hline
\end{tabular}
\end{minipage}
\end{table*}

\begin{table*}
 \centering	
 \begin{minipage}{119mm}
 \contcaption{}
  \begin{tabular}{@{}rrrrcccrrr@{}}
  \hline
   RA$_{pri}$ & Dec$_{pri}$ & RA$_{sat}$ & Dec$_{sat}$ & M$_{B(pri)}$ & M$_{B(sat)}$ & inc$_{pri}$ & $r_p$ & $\theta$ & $v_r$\\
 \hline
166.496 & 58.946 & 166.640 & 58.922 & -20.4 & -17.7 & 51 & 264 & 26 & 83 \\
173.701 & 46.990 & 173.739 & 46.889 & -20.0 & -17.0 & 63 & 248 & 63 & 196 \\
173.701 & 46.990 & 173.781 & 46.770 & -20.0 & -17.6 & 63 & 535 & 64 & -177 \\
173.853 & 57.650 & 173.849 & 57.650 & -19.9 & -17.0 & 37 & 5 & 15 & 131 \\
173.853 & 57.650 & 173.792 & 57.579 & -19.9 & -17.5 & 37 & 162 & 79 & -462 \\
174.028 & 62.249 & 173.932 & 62.258 & -20.1 & -17.6 & 60 & 107 & 86 & 205 \\
174.028 & 62.249 & 173.433 & 62.361 & -20.1 & -17.0 & 60 & 702 & 75 & 29 \\
181.347 & 64.508 & 181.220 & 64.605 & -21.5 & -18.9 & 35 & 618 & 12 & 64 \\
191.340 & 61.694 & 191.596 & 61.781 & -20.7 & -18.7 & 34 & 606 & 62 & 493 \\
192.416 & 49.447 & 192.273 & 49.554 & -20.6 & -18.0 & 43 & 390 & 16 & 199 \\
193.704 & 44.156 & 193.537 & 44.026 & -20.7 & -18.9 & 70 & 672 & 63 & -4 \\
198.250 & 43.204 & 198.285 & 43.217 & -20.9 & -18.8 & 40 & 121 & 79 & -64 \\
203.203 & 41.872 & 203.148 & 41.788 & -20.3 & -16.7 & 31 & 182 & 75 & -86 \\
203.203 & 41.872 & 203.341 & 42.146 & -20.3 & -17.5 & 31 & 573 & 70 & -37 \\
204.542 & 60.273 & 204.221 & 60.329 & -20.3 & -17.0 & 43 & 406 & 0 & -128 \\
204.542 & 60.273 & 204.841 & 60.293 & -20.3 & -17.6 & 43 & 361 & 27 & -183 \\
207.873 & 43.806 & 207.891 & 43.820 & -19.9 & -16.7 & 47 & 44 & 63 & 222 \\
220.880 & 49.393 & 221.081 & 49.412 & -19.6 & -17.4 & 30 & 287 & 79 & 47 \\
221.826 & 58.226 & 221.557 & 58.004 & -18.7 & -17.0 & 55 & 707 & 10 & -70 \\
222.111 & 34.998 & 222.108 & 35.159 & -20.3 & -16.8 & 69 & 341 & 41 & 5 \\
222.912 & 40.599 & 222.731 & 40.524 & -19.7 & -15.2 & 61 & 187 & 23 & -1 \\
222.912 & 40.599 & 222.507 & 40.362 & -19.7 & -15.1 & 61 & 464 & 14 & 67 \\
226.179 & 61.719 & 226.284 & 61.703 & -18.7 & -16.4 & 49 & 107 & 8 & 217 \\
229.855 & 45.880 & 229.831 & 45.790 & -19.3 & -17.2 & 71 & 115 & 71 & 40 \\
238.795 & 52.169 & 238.716 & 52.134 & -20.0 & -17.9 & 50 & 160 & 44 & -91 \\
240.751 & 27.010 & 240.649 & 27.212 & -20.5 & -17.1 & 35 & 517 & 19 & 149 \\
240.751 & 27.010 & 240.684 & 27.201 & -20.5 & -16.5 & 35 & 466 & 12 & 343 \\
241.370 & 42.627 & 241.180 & 42.712 & -20.5 & -17.5 & 49 & 459 & 60 & -126 \\
242.683 & 41.149 & 242.797 & 40.944 & -20.0 & -16.4 & 66 & 504 & 14 & -288 \\
355.895 & 0.568 & 355.903 & 0.635 & -21.0 & -18.7 & 54 & 376 & 7 & 114 \\
\hline
\end{tabular}
\end{minipage}
\end{table*}

\begin{table*}
 \centering
 \begin{minipage}{64mm}
  \caption{Bok-Observed Primaries with No Detected Rotational Sense.  
  Same units as Table 1.  Apparent magnitudes
are AB SDSS Petrosian r-band.}
  \begin{tabular}{@{}rrccc@{}}
  \hline
  RA$_{pri}$ & Dec$_{pri}$ & m$_{r(pri)}$ & M$_{B(pri)}$ & inc$_{pri}$\\
 \hline
27.938 & $-$9.240 & 15.3 & $-$20.7 & 41 \\
113.723 & 39.764 & 15.1 & $-$20.8 & 35 \\
123.097 & 28.407 & 15.2 & $-$20.5 & 28 \\
127.257 & 47.801 & 14.4 & $-$20.3 & 32 \\
128.955 & 51.583 & 15.1 & $-$19.5 & 66 \\
134.424 & 43.127 & 13.9 & $-$20.2 & 55 \\
141.913 & 7.080 & 14.9 & $-$20.8 & 56 \\
146.196 & 51.689 & 14.3 & $-$20.0 & 29 \\
148.870 & 38.641 & 14.8 & $-$20.7 & 38 \\
148.996 & 6.566 & 15.1 & $-$20.6 & 52 \\
152.939 & 11.346 & 15.0 & $-$20.4 & 38 \\
156.349 & 64.999 & 14.8 & $-$19.0 & 51 \\
157.337 & 7.768 & 15.3 & $-$19.0 & 44 \\
157.583 & 8.061 & 14.6 & $-$20.4 & 43 \\
158.009 & 13.774 & 15.4 & $-$21.3 & 63 \\
161.873 & 7.251 & 13.7 & $-$19.9 & 42 \\
162.751 & 12.287 & 14.3 & $-$20.2 & 41 \\
164.687 & 59.511 & 13.5 & $-$19.8 & 38 \\
165.657 & 59.125 & 13.7 & $-$20.4 & 67 \\
169.969 & 54.463 & 14.3 & $-$20.5 & 29 \\
171.812 & 10.322 & 14.4 & $-$20.6 & 54 \\
173.097 & 54.983 & 14.8 & $-$20.1 & 51 \\
173.950 & 63.492 & 14.4 & $-$19.6 & 54 \\
176.858 & 54.407 & 15.3 & $-$20.7 & 54 \\
183.797 & 54.634 & 15.2 & $-$21.3 & 33 \\
184.344 & 53.557 & 14.7 & $-$20.7 & 50 \\
184.649 & 42.014 & 14.7 & $-$20.4 & 57 \\
188.135 & 48.472 & 14.6 & $-$20.9 & 53 \\
189.000 & 54.221 & 13.1 & $-$19.4 & 40 \\
189.306 & 49.448 & 13.7 & $-$20.2 & 47 \\
189.661 & 1.473 & 14.6 & $-$20.3 & 66 \\
190.146 & 2.468 & 14.4 & $-$20.2 & 56 \\
193.903 & 60.183 & 15.2 & $-$20.8 & 65 \\
194.925 & 43.753 & 14.6 & $-$20.8 & 52 \\
197.675 & 42.285 & 14.6 & $-$19.9 & 48 \\
200.872 & 49.013 & 15.2 & $-$18.6 & 43 \\
201.021 & 58.287 & 15.2 & $-$19.7 & 48 \\
204.874 & 46.564 & 14.0 & $-$20.3 & 35 \\
210.105 & 65.218 & 15.1 & $-$20.9 & 50 \\
210.689 & 38.066 & 14.9 & $-$20.8 & 54 \\
211.579 & 36.834 & 15.3 & $-$19.8 & 69 \\
222.210 & 57.771 & 14.5 & $-$20.1 & 33 \\
222.327 & 52.553 & 15.1 & $-$20.7 & 46 \\
223.321 & 52.044 & 14.4 & $-$20.1 & 42 \\
224.259 & 35.551 & 14.8 & $-$21.1 & 45 \\
226.154 & 48.739 & 14.9 & $-$19.4 & 34 \\
227.527 & 56.373 & 14.7 & $-$20.4 & 52 \\
229.589 & 58.112 & 14.6 & $-$19.3 & 64 \\
231.464 & 38.771 & 15.0 & $-$20.2 & 40 \\
232.739 & 36.807 & 14.8 & $-$20.2 & 67 \\
232.828 & 36.080 & 15.1 & $-$19.2 & 32 \\
232.952 & 51.768 & 14.5 & $-$18.6 & 43 \\
235.291 & $-$1.706 & 14.0 & $-$20.4 & 45 \\
\hline
\end{tabular}
\end{minipage}
\end{table*}

\begin{table*}
\centering
\begin{minipage}{64mm}
\contcaption{}
\begin{tabular}{@{}rrccc@{}}
\hline
  RA$_{pri}$ & Dec$_{pri}$ & m$_{r(pri)}$ & M$_{B(pri)}$ & inc$_{pri}$\\
\hline
241.073 & 42.028 & 15.0 & $-$19.6 & 51 \\
241.823 & 41.404 & 14.5 & $-$19.8 & 36 \\
241.907 & 45.065 & 14.5 & $-$20.0 & 57 \\
253.593 & 41.335 & 13.7 & $-$20.2 & 33 \\
258.118 & 30.910 & 14.5 & $-$20.9 & 39 \\
323.174 & $-$0.128 & 14.4 & $-$20.9 & 56 \\
332.970 & 0.109 & 14.0 & $-$20.2 & 46 \\
335.206 & $-$9.488 & 15.4 & $-$20.9 & 39 \\
349.606 & 14.939 & 15.3 & $-$19.7 & 52 \\
357.035 & $-$10.774 & 15.5 & $-$20.4 & 51 \\
\hline
\end{tabular}
\end{minipage}
\end{table*}

\bsp
\label{lastpage}
\end{document}